\documentclass[aps,prl,twocolumn,amssymb,amsfonts,groupedaddress,showpacs,epsf]{revtex4}
\usepackage{epsfig}
\usepackage{amsmath}
\usepackage{bm}
\usepackage{times}

\begin{document}
\title{Electrically controlled quantum gates for two-spin qubits in two double quantum dots}

\author{Guy Ramon}
\email{gramon@scu.edu}
\affiliation{Department of Physics, Santa Clara University, Santa Clara, CA 95053}

\begin{abstract}

Exchange-coupled singlet-triplet spin qubits in two gate-defined double quantum dots are considered theoretically.
Using charge density operators to describe the double-dot orbital states, we calculate the Coulomb couplings between the qubits, and identify optimal bias points for single- and two-qubit operations, as well as convenient idle positions. The same intuitive formulation is used to derive dephasing rates of these qubits due to the fluctuating charge environment, thereby providing the main considerations for a quantum computation architecture that is within current experimental capabilities.

\end{abstract}


\maketitle

Electron spins localized in semiconductor quantum dots (QDs) have been vigorously pursued in recent years as potential qubit candidates, due to their relative isolation from the environment \cite{HanRMP}. The long coherence times measured for single electron spins render their manipulation and readout particularly challenging. This has prompted a number of proposals to encode the logical qubit into two-spin singlet ($S$) and unpolarized triplet ($T_0$) states \cite{Lidar,Levy,Taylor}.
GaAs gate-defined lateral QDs in particular have served as a useful platform to study quantum control and decoherence of exchange-coupled spin qubits. The main reason is the high tunability of the two-electron exchange interaction, $J$, that is controlled on a subnanosecond scale by changing the bias between the two dots. In the context of the $S-T_0$ qubit this tunability enables initialization of the qubit state in a doubly occupied singlet state using positive interdot bias, where triplet states are energetically inaccessible. By applying a perpendicular magnetic field, the two polarized triplet states are Zeeman splitted, further reducing leakages from the logical basis states. Recent experiments in gated double dots have implemented spin-echo \cite{Petta,Barthel,Bluhm} and pulse optimization techniques that have pushed two-spin qubit dephasing time over $200 \mu s$ \cite{Bluhm}.

\begin{figure}[tb]
\epsfxsize=0.95\columnwidth
\vspace*{0.6 cm}
\centerline{\epsffile{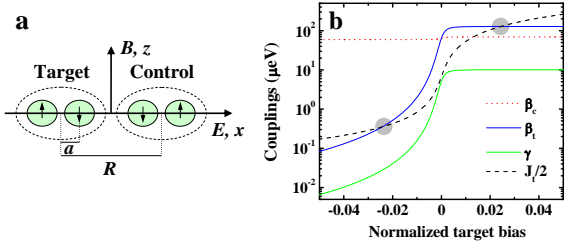}}
\vspace*{-3.6 cm}
\caption{(color online) (a) Geometry of the two double-dot qubits. We denote the left qubit as the target, and the right qubit as the control; (b) Two-qubit Coulomb couplings, and exchange interaction vs.~target qubit bias $\epsilon_t$, for a fixed control qubit bias $\epsilon_c=0$ and inter-qubit distance $R=300$ nm. Locations of zero effective exchange are marked by Gray circles. Inter-dot bias is normalized to QD confinement and detuning is measured from the singlet anticrossing point. }
\label{Fig1}
\end{figure}

In this work we consider two proximal $S-T_0$ qubits, each formed by two electrons localized in a gate-defined double dot, as depicted in Fig.~\ref{Fig1}(a) \cite{Taylor,Stepanenko}. Single qubit $Z$-rotations are fixed by pulsing the interdot bias that controls $J$, whereas $X$-rotations are generated by a magnetic field gradient across the two dots. A promising method to produce such gradient is to utilize the hyperfine coupling at the $S-T_+$ degeneracy point to dynamically induce unequal nuclear polarization across the two dots. Implementing these electrically controlled pump cycles, Foletti {\it et al.} were recently able to sustain a stable Overhauser field gradient of $\delta h \sim 5 \mu$eV \cite{Foletti,BluFol}. The single-qubit Hamiltonian can thus be written as ${\cal H}_q={\bf B} \cdot \mbox{\boldmath $\sigma$}$, where ${\bf B}=\frac{1}{2}(\delta h,0,J)$, and $\mbox{\boldmath $\sigma$}$ is the vector of Pauli spin matrices for the pseudospin states $S$, and $T_0$.

When two double dots are placed near each other, their respective electrons become Coulomb coupled. This coupling depends on the orbital states and thus on the spin configurations of the electrons, resulting in an effective additional exchange interaction for each qubit, and spin dependent entangling coupling. Below we show that for certain bias points the first term turns off the exchange interaction that is otherwise constantly on. The latter term provides a controlled-phase (CPHASE) gate, thus completing a universal set of gates that is based solely on accurate control over the interdot bias at each qubit. In particular, the requirement to control the tunnel coupling between the dots, needed to tune $J$ to zero \cite{Taylor,Imp}, and the need to perform single-qubit $Z$-rotations near the dephasing-susceptible avoided crossing point \cite{HanBur}, are eliminated, and qubit decoupling positions are identified.

\emph{Inter-qubit Coulomb couplings and sweet spots.} The Coulomb interaction between two qubits, each formed by the singlet and triplet spin states of two electrons in a biased double dot, is found by considering the charge distributions of the electron orbital states, assuming no inter-qubit tunnel coupling. Up to a constant term, the interaction Hamiltonian can be written in the form \cite{Ramon}:
\begin{equation}
{\cal H}_{\rm int}=-\beta_t \sigma^z_t -\beta_c \sigma^z_c -\gamma \sigma^z_t \otimes \sigma^z_c,
\label{Hint}
\end{equation}
where
\begin{eqnarray}
\beta_{t/c} &=& \frac{1}{4} \left( V_{{\rm T_0 T_0}}\mp V_{{\rm ST_0}}\pm V_{{\rm T_0 S}}-V_{{\rm SS}} \right) \nonumber\\
\gamma &=& \frac{1}{4} \left(-V_{\rm T_0 T_0}+V_{\rm ST_0}+V_{\rm T_0 S}-V_{\rm SS} \right).
\label{bg}
\end{eqnarray}
Here the subscript $t$ ($c$) denotes target (control) qubit, and $V_{ij}$ are the Coulomb matrix elements involving the hybridized singlet ($S$) and unpolarized triplet ($T_0$) states, where the left (right) subscript corresponds the target (control) qubit. For nominally identical, equally biased QDs, we have $\beta_t=\beta_c$. When the inter-qubit distance $R$ is sufficiently large, there is no tunneling between the qubits, the two charge distributions are separated in space, and the Coulomb interaction between them can be described systematically using a multipole expansion approach \cite{Ramon,Aux}. For the same-axis geometry shown in Fig.~\ref{Fig1}(a), couplings are always positive, and they are plotted in Fig.~\ref{Fig1}(b) as a function of the target qubit bias, $\epsilon_t$, for $R=300$ nm and a fixed control qubit bias, $\epsilon_c$, corresponding to the anticrossing point. Here and for all results reported in this work we have considered $B=100$ mT ($E_Z=2.5 \mu {\rm eV}$), and have modeled the double dots using a quartic potential with dot confinement $\omega_0=3 {\rm meV}$ ($a_B\approx 20$ nm), and half interdot distance $a=2.8a_B$. While all the Coulomb couplings depend on both $(\epsilon_t,\epsilon_c)$, for the parameter range considered, each of the $\beta$ couplings is predominantly controlled by its respective qubit bias (as seen by the red dotted line in Fig.~\ref{Fig1}(b)).

\begin{figure}[tb]
\epsfxsize=0.7\columnwidth
\vspace*{0 cm}
\centerline{\epsffile{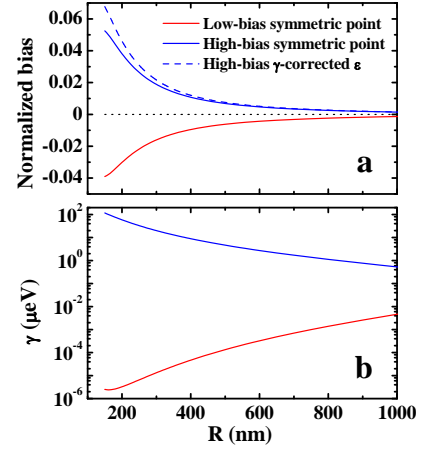}}
\vspace*{-0.2 cm}
\caption{(color online) (a) Normalized target and control qubit bias at symmetric low-bias (solid red line) and high-bias (solid blue line) points vs.~$R$. The dashed blue line depicts the high-bias position for the $\gamma$-corrected sweet-spot. The dotted black line marks the $S(1,1)-S(0,2)$ anticrossing bias point; (b) $\gamma$ coupling values at low-bias (red line) and high-bias (blue line) sweet spots.}
\label{Fig2}
\end{figure}

Complementing the target and control single-qubit Hamiltonians with the interaction terms, Eq.~(\ref{Hint}), we find that at certain bias points exchange is turned off (gray circles in Fig.~\ref{Fig1}b). We define a sweet spot in the 2D bias space $(\epsilon_t,\epsilon_c)$ as a point that satisfies two conditions: ${\tilde J}_t(\epsilon_t,\epsilon_c)= J_t-2\beta_t=0$, and ${\tilde J}_c(\epsilon_t,\epsilon_c)= J_c-2\beta_c=0$. The first condition facilitates an idle position of the target qubit, or a pure $X$-rotation in the presence of a magnetic field gradient, $\delta h_t$. Similarly, the second condition is set to turn off the control qubit exchange, resulting in minimal interference with the target operation. For $R\lesssim 1200$ nm we find that for each qubit there are always two such intersections, thus in the 2D bias space there are four possible working positions: two symmetric low- and high-bias points, and two asymmetric points. As explained below, the symmetric low- and high-bias points are optimal for single- and two-qubit operations, respectively. Fig.~\ref{Fig2}(a) depicts the dependence of the symmetric low-bias (solid red line) and  high-bias (solid blue line) sweet-spot positions on the inter-qubit distance $R$. The corresponding entangling coupling, $\gamma$, at these bias positions is shown in Fig.~\ref{Fig2}(b). For almost the entire range, the low-bias $\gamma$ values exhibit a counter-intuitive increase with $R$ due to the increased sweet-spot bias as $R$ grows. The strong bias dependence of $\gamma$ overtakes its expected decrease with $R$ for the considered range.

\emph{Idle position and single-qubit gates.} While the $\gamma$ coupling cannot induce relaxation of the two-electron qubit states, it generates dephasing of a superposition $S$-$T_0$ state, and gate errors. Inspecting the two-qubit Hamiltonian, we find that without magnetic field gradients, $\delta h_t =\delta h_c =0$, when the two qubits are biased at a sweet-spot, preparing one of them in the state $(|S \rangle+|T_0 \rangle )/\sqrt{2}$ results in a time evolution of its off-diagonal density matrix element: $\rho_{ST_0}(t) = \frac{1}{2} \cos 2\gamma t$, irrespective of the state of the other. Choosing the low-bias sweet-spot and $R=300$ nm, we have $\gamma=0.013$ neV, corresponding to dephasing time of $\sim 40 \mu$s.

Since nuclear state preparation that generates a stable magnetic field gradient requires pumping times between 60 ms - 1 s \cite{Foletti}, it may well be necessary to operate under a fixed $\delta h$ in both qubits. As long as $\delta h_c >> \gamma$, the oscillation amplitude of the off diagonal element of the target qubit density matrix reads $2 (\gamma/\delta h_c)^2$. Taking a conservative value of $\delta h_c=1 \mu$eV, corresponding to a field gradient of $\sim 40$ mT, we find that at $R=300$ nm, target qubit errors are below $10^{-9}$. Working with a finite $\delta h_t=\delta h_c =\delta h$, the sweet-spot is no longer an idle position for the target qubit, and the gate time for a $\pi$-flip $X$-rotation is 2 ns, with a leading gate error of $(\pi\gamma/2\delta h)^2 \lesssim 4\times 10^{-10}$, for $\delta h=1 \mu eV$ \cite{Aux}. If the $\tilde{J}_t=0$ point is used as an idle position, $X$-rotations can be erased either by waiting integer number of full periods, $\tau=2 \pi n/\delta h_t$, or by applying the pulse sequence: $U_Z(\pi) U_X(\alpha)U_Z(\pi)$, where $\alpha =\delta h_t \tau$ is the acquired $X$-rotation during the wait time $\tau$.

To generate $Z$-rotations with a finite $\delta h$ one can positively bias the target qubit so that $\tilde{J}_t >>\delta h$ produces a nearly perfect $Z$-rotation. Note, however, that to maintain short gate times, $\delta h$ cannot be too small, thus a large excursion from the sweet-spot is necessary to obtain sufficiently large $\tilde{J}_t$. At this position, near the singlet anticrossing, the $\gamma$ coupling increases dramatically, generating substantially larger gate errors. In addition, a positively biased qubit may decohere faster due to environmental charge fluctuations. To stay near the $\tilde{J}_t=0$ point, one can perform a three pulse sequence: $U_z (\alpha)=U_\theta (\chi) U_X (\phi) U_\theta (\chi)$, where $U_\theta (\chi)$ is a rotation about the $\theta$-tilted axis, $\theta=\tan^{-1} (\delta h/\tilde{J}_t)$, and the rotation angles $\chi$ and $\phi$ are fixed by $\alpha$ and $\theta$ \cite{HanBur,Aux}. A $Z$-rotation by any angle $\alpha$ can be performed using this cycle, as long as $|\tilde{J}_t| \geq \delta h$ ($|\theta| \leq \pi/4$). Choosing $\theta=\pi/4$ to stay as close as possible to the $\tilde{J}_t=0$ point, we complete a three-pulse $Z$-rotation of $\alpha=\pi$ in 5 ns, with a gate error smaller than $10^{-7}$. The dominant contribution to the $Z$-gate error reads: $2(\gamma_\theta/\delta h)^2(3-\cos \sqrt{2} \pi)$, where $\gamma_\theta$ is evaluated at the $\theta$-tilted bias position. Although $\gamma_\theta$ increases with $\delta h$ due to the necessary larger bias sweep to the $\tilde{J}_t=\delta h$ position, the ratio $\gamma_\theta/\delta h$ is monotonically decreasing in the considered range, keeping the gate error very small for $\delta h \lesssim 10 \mu eV$ \cite{Aux}.

\emph{Two-qubit gates.} We now consider the high bias symmetric point, where the qubit coupling $\gamma$ is large: $\gamma \gg \delta h$. Without magnetic field gradients, the two-qubit Hamiltonian is diagonal in the computational basis state $\{SS,T_0 S,ST_0,T_0 T_0\}$, and reads at the sweet-spot: ${\cal H}={\rm diag} (-\gamma,\gamma,\gamma,-\gamma)$. A CPHASE gate is conveniently obtained by tuning the bias in both double dots to satisfy $J_q=2(\beta_q+\gamma)$, $q=t,c$, at which ${\cal H}={\rm diag}(\gamma, \gamma, \gamma, -3\gamma)$.
Letting the system evolve for time $\tau_{\rm cp}=\pi \hbar/4 \gamma$, the resulting gate, up to a common phase, is: ${\rm diag} (1,1,1,-1)={\rm CPHASE}$. The position of the $\gamma$-corrected high-bias symmetric point is shown by the dashed line in Fig.~\ref{Fig2}(a).

\begin{figure}[tb]
\epsfxsize=0.95\columnwidth
\vspace*{2.5 cm}
\centerline{\hspace{0.5 cm} \epsffile{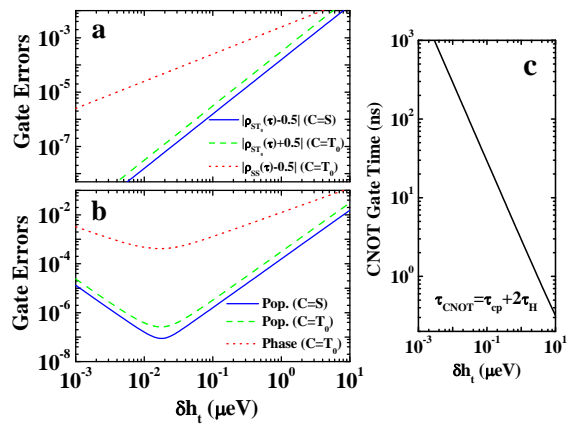}}
\vspace*{-2.7 cm}
\caption{(color online) (a) CPHASE gate errors vs. $\delta h$, for a target qubit initially at $(| S \rangle -|T_0\rangle)/\sqrt{2}$, and $R=300$ nm. The solid blue (dashed green) line depicts phase error $\sim (\delta h/2\gamma)^2$ for control in singlet (triplet) state, and the dotted red line shows population errors $\sim \delta h/2\gamma$, when control is in triplet. Population errors when control is in singlet are very small (not shown); (b) CNOT gate errors vs. $\delta h_t$, for an initial target singlet state (same results are obtained for an initial target triplet state). The solid blue (dashed green) line depicts population errors for control in singlet (triplet) state, and the dotted red line shows phase error when control is in triplet. Phase error when control is in singlet is very small (not shown); (c) CNOT gate time vs.~$\delta h_t$. Note that $\tau_{\rm cp}=0.026$ ns is fixed by $\gamma$, and most of the CNOT gate time is spent on the Hadamard target gates.}
\label{Fig3}
\end{figure}

Working with finite field gradients in both qubits, we find three different gate error scalings: (i) population errors with control in $|T_0 \rangle$ $\sim \delta h/2\gamma$, (ii) phase errors $\sim (\delta h/2\gamma)^2$, and (iii) population errors  with control in $|S \rangle$ $\sim (\delta h/2\gamma)^3$, the latter being negligible in the considered $\delta h$ range. We have performed numerical simulations of the time evolution of the two qubits under a controlled $\pi$-pulse, where the target qubit is initially prepared in the state: $(| S \rangle +|T_0\rangle)/\sqrt{2}$, and the control qubit is in either $|S\rangle$ (target remains in its initial state) or $|T_0 \rangle$ (target evolves to $(| S \rangle -|T_0\rangle)/\sqrt{2}$). Fig.~\ref{Fig3}(a) shows the resulting CPHASE gate errors for $R=300$ nm ($\gamma =20 \mu eV$), which follow closely the above scalings up to $\delta h \lesssim 10 \mu eV$. The small difference in the two phase errors with different control qubit states is due to the Hamiltonian structure at the $\gamma$-corrected working point.

To construct a CNOT gate, a Hadamard gate, $U_H=(U_X(\pi)+U_Z(\pi))/\sqrt{2}$, should be applied to the target qubit before and after the controlled operation \cite{CNOT}. $U_H$ can be implemented by biasing the target qubit to a position adjacent to the low-bias sweet spot, where $\tilde{J}_t=\delta h_t$, for a duration $\tau_H =\pi \hbar/ \sqrt{\tilde{J}_t^2+\delta h_t^2}$. We note that the control qubit should not rotate prior to the CPHASE gate. While it may be possible to employ a pulse sequence to correct such spurious control rotation, here we take $\delta h_c=0$ during the first Hadamard target gate. This may be implemented by, e.g., delaying the nuclear state preparation in the control qubit by $\tau_H$. This pulse sequence was simulated by discretizing bias and time steps, so that the actual switching times between different bias points are taken into account. Fig.~\ref{Fig3}(b) shows the resulting CNOT gate errors vs.~$\delta h_t$ for $R=300$ nm. Population and phase errors interchange with respect to their role in the CPHASE gate, therefore the phase error with control in $|T_0\rangle$ is the largest here, scaling linearly with $\delta h_t/2\gamma$. The results demonstrate increased errors for both small $\delta h_t$ (during the Hadamard gates), and large $\delta h_t$ (during the CPHASE gate). A minimum phase error of $4 \times 10^{-4}$ is obtained for $\delta h_t=0.017 \mu eV$, but with such a small gradient, single qubit gate times are on the order of 100 ns (see Fig.~\ref{Fig3}(c)). Phase errors smaller than $10^{-3}$ are obtained with $0.004 \lesssim \delta h_t \lesssim 0.07 \mu eV$. Other errors, and in particular singlet return errors, are considerably smaller.

We stress that the opposing requirements for small single- and two-qubit gate errors, while operating under fixed $\delta h$, are met by utilizing the high tunability of the $\gamma$ coupling, whose values show more than six orders-of-magnitude difference between the low- and high-bias sweet-spots for $R=300$ nm. Taking $R=200$ nm will increase the $\gamma$ ratio by another order-of-magnitude, resulting in a minimum phase error of $6 \times 10^{-5}$ at $\delta h_t=0.02 \mu eV$, and a wider range for which errors are below $10^{-3}$ up to $\delta h_t \lesssim 0.65 \mu eV$. Reducing the inter-qubit distance may, however, induce errors due to electron exchange coupling between the two double dots that are not included in the current analysis. Further optimization of gate errors and times is likely to be obtained by engineering the dot size and tunnel coupling within each double dot \cite{HanBur}.

\emph{Qubit dephasing due to fluctuating charge environment.} Exchange-coupled spin qubits are vulnerable to dephasing induced by charge noise, due to the different charge distributions of the singlet and triplet states \cite{Hu,Culcer,Ramon}. The formulation we employed to find inter-qubit couplings can be used to quantify couplings of each of the qubits to nearby two-level charge fluctuators (TLFs) and evaluate the resulting qubit dephasing. We identify two types of TLFs: (i) $\beta$-coupled, where the charge jumps in and out of the trap, and (ii) $\gamma$-coupled, where the charge fluctuates between two sites in the trap. While $\beta$-coupled TLFs have stronger coupling to the spin qubits, they are less abundant and locate farther from the QDs, near the 2DEG layer or quantum point contacts.

Starting with a single TLF, we represent the qubit state with a Bloch vector, and solve for its precession under the pseudo-field: ${\bf B}(t)=(\delta h,0,\tilde{J}-v \xi(t))$ \cite{Bergli}. $\xi(t)=\pm 1$ represents a Poisson switching process, and $v=\beta, 2\gamma$ is the qubit-TLF coupling strength for $\beta$-type and $\gamma$-type TLFs, respectively. Averaging over the stochastic process, the qubit signal dephasing is given by: $W(t)=\langle e^{i\varphi t} \rangle$, with $\varphi = v \int_0^t \xi(t') dt'$. For free induction decay (FID), we find \cite{Paladino}:
\begin{equation}
W_{\rm FID} (t)= \frac{e^{-\Gamma t}}{4\mu} \sum_\pm \pm \left[ (1 \pm \mu)^2 +\left(\frac{v}{\Gamma} \right)^2 \right] e^{\pm\Gamma \mu t},
\label{fid}
\end{equation}
where $\mu^2=1-(v/\Gamma)^2+2iv/\Gamma \tanh (\Delta E/2k_B T)$, $\Gamma=\frac{1}{2}(\Gamma_+ +\Gamma_-)$ is the average switching rate, $\Gamma_+/\Gamma_- =e^{-\Delta E/k_B T}$, and $\Delta E$ is the TLF level splitting. To calculate signal decay during a general dynamical decoupling pulse sequence we write $\varphi=v \int_0^t \zeta (t,t') \xi (t') dt'$, where $\zeta (t,t')$ is a filter function in the time domain \cite{Cywinski}. For the spin echo (SE) protocol, $\zeta(t,t')=\Theta(t')\Theta(t/2-t')-\Theta(t'-t/2)\Theta(t-t')$, where $\Theta (t)$ is the step function, and the signal decay is found to be:
\begin{eqnarray}
W_{\rm SE} (t)\!&\!=\!&\! \frac{e^{-\Gamma t}}{2|\mu|} \left[ (\mu_I^2+1) \sum_\pm (1\pm \mu_R) e^{\pm \Gamma \mu_R t} \right. \nonumber \\
\!&\!+\!&\! \left. (\mu_R^2-1) \sum_\pm (1 \pm i\mu_I) e^{\pm i\Gamma \mu_I t} \right],
\label{se}
\end{eqnarray}
where $\mu_R$ ($\mu_I$) is the real (imaginary) part of $\mu$.

In order to estimate two-spin qubit dephasing due to an ensemble of TLFs, we use $W(t)=\prod_{i=1}^{n_\beta} W_i^\beta (t) \prod_{i=1}^{n_\gamma} W_i^\gamma (t)$, where $n_\beta$ ($n_\gamma$) is the number of $\beta$-type ($\gamma$-type) traps. We estimate the trap density as $\sim 120 \mu m^{-2}$, and the two double-dot device area as $\sim 0.75 \mu m^2$, resulting in an active number of traps: $n_\beta =3$, $n_\gamma =9$  \cite{Traps}. In this estimate we assume that an area of radius $\sim 150$ nm around each qubit is depleted, so that traps in this area are not charge-active, setting up the maximum qubit-TLF coupling strength. Trap parameters include level splitting of $10 \mu eV$, center radius of 5 nm, and intersite distance of 20 nm, chosen to characterize $\delta$-doped dopants in the insulator. Finally, trap switching rates are taken between 1ms $< \Gamma^{-1} < 1$ s \cite{Pioro}.

We have used Eqs.~(\ref{fid}) and (\ref{se}) to calculate dephasing of a qubit at the low-bias sweet spot, by averaging over 10,000 sets of TLFs with random locations, orientations, and switching rates, within the parameter ranges specified above. In the FID case we obtain dephasing time of $T_2^*=23$ ns, comparable to measured coherence times \cite{Petta}. For the SE protocol we find $T_2^{\rm SE}=162 \mu s$, two-orders-of-magnitude longer than measured SE coherence times \cite{Bluhm}, attributed to nuclear spin fluctuations, mediated by the hyperfine interaction \cite{Cywinski,Neder}. In the quasi-static regime considered here, FID dephasing times are dominated by the strongest fluctuator(s), and are thus mostly sensitive to qubit bias and qubit-TLS distance. SE dephasing times are dominated by the fastest fluctuator(s), and are sensitive to the maximal switching rate \cite{Ramon2}. We stress that for such a mesoscopic system, these results are not self averaging and may vary considerably between samples. For SE experiments with larger devices, sample-to-sample variability is reduced and dephasing times become comparable with those induced by the nuclei \cite{Aux}, thus both charge and nuclear environments should be considered in future device design and operation.

In conclusion, we have quantified the couplings between two $S-T_0$ qubits, identifying optimal working points. Our proposed scheme relies solely on interdot bias tuning at each qubit, and offers several advantages, including high fidelity single- and two-qubit gates under fixed field gradients, and idle positions at which the qubits are effectively decoupled. In addition, we have used our formulation to quantify qubit dephasing due to charge fluctuations.

The author thanks Xuedong Hu and Hendrik Bluhm for helpful discussions and acknowledges funding from Research Corporation.

\bibliographystyle{amsplain}

\begin{thebibliography}{99}
\bibitem{Imp} Fast control over the tunnel coupling in gate defined double dots is challenging and has not been demonstrated yet. In other QD systems such as self-assembled or vertically stacked QDs tunnel coupling is fixed by fabrication.
\bibitem{CNOT} Alternative implementations of CNOT gate are briefly discussed in \cite{Aux}.
\end{thebibliography}

\clearpage

\renewcommand{\theequation}{S\arabic{equation}}
\setcounter{equation}{0}
\renewcommand{\thefigure}{S\arabic{figure}}
\setcounter{figure}{0}

\section*{Supplemental material for: Electrically controlled quantum gates for two-spin qubits in two double quantum dots}

\subsection*{Coulomb couplings between two double-dots}

Here we outline the calculation of the Coulomb couplings between the two-electron orbital states associated with each qubit. Since all the gate operations considered in this work are performed by tuning the bias between the two QDs, the Hund-Mulliken orbital Hamiltonian must account for biased dot configuration. More details are given in \cite{Ramon}.

We start by approximating the orbitals for the biased double-dot by those of two harmonic wells centered at $(\pm a +\epsilon a_B)\hat{x}$, where $a$ is the interdot half separation, $a_B$ is the QD Bohr radius, and $\epsilon=e E a_B/\hbar \omega_0$ is a dimensionless parameter corresponding to the interdot bias (normalized to the QD confinement). We denote the orthonormalized (predominantly) left and right single particle orbitals by $\psi_{\pm a}$, and use them to construct the two-electron spin basis states. These are the two doubly occupied singlet states, $S(2,0)=\psi_{-a} \psi_{-a}$, $S(0,2)=\psi_{a} \psi_{a}$, and the separated singlet and unpolarized triplet states, $S(1,1)=(\psi_{-a} \psi_{a}+\psi_{a} \psi_{-a})/\sqrt{2}$, $T_0(1,1)=(\psi_{-a} \psi_{a}-\psi_{a} \psi_{-a})/\sqrt{2}$, where$(i,j)$ indicates the number of electrons in each dot. Since the orbital Hamiltonian does not connect the triplet state with any of the singlet states, we only need to diagonalize the $3 \times 3$ singlet block, resulting in a hybridized singlet state.

We note that both the polarized and doubly occupied triplet states are neglected in our analysis. Doubly occupied triplet states require one of the electrons to reside in an excited orbital, thus their energy is far above the doubly occupied singlets \cite{Petta}. Utilizing Overhauser fields to generate a magnetic field gradient will likely create a small perpendicular component that may mix the $S_z=0$ computational subspace with the $S_z= \pm 1$ subspace , spanned by the polarized triplet states. With $\delta h =1\mu eV$, the transverse inhomogeneous field should not exceed $\delta h_\perp = 50$ neV for which leakage errors are found to be below $10^{-5}$ \cite{HanBur}, thus polarized triplets can be safely neglected as well.

For the Calculation of the inter-qubit Coulomb couplings, it is convenient to work in the orthonormal basis of symmetric and antisymmetric orbital combinations: $\psi_1=(\psi_{-a}+\psi_a)/\sqrt{2}$, $\psi_2=(\psi_{-a}-\psi_a)/\sqrt{2}$. In this basis, the Coulomb interaction operator between the two qubits is written as
\begin{equation}
f_{ijkl}=\int d{\bf r} d{\bf r}' \frac{\rho^t_{ij}({\bf r}) \rho^c_{kl}({\bf r}')}{\varepsilon |{\bf r}-{\bf r}'|}
\end{equation}
where $\rho_{ij}({\bf r})=e\psi_i^*({\bf r}) \psi_j ({\bf r})$ is the electron charge density operator for either qubit, and $i,j \in \{1,2\}$ denote the qubit orbital state (symmetric or antisymmetric combination). The Coulomb matrix elements in Eq.~(\ref{bg}) in the main text are found in terms of the $f_{ijkl}$ operators:
\begin{eqnarray}
V_{T_0 T_0} \!&\! =\!&\! f_{1111}+f_{1122}+f_{2211}+f_{2222} \nonumber \\
V_{T_0 S} \!&\!=\!&\! V_{T_0 T_0}+{\cal N}_c^2 \left( a_1^c+a_2^c \right)\left[ \sqrt{2} \left( f_{1111}+f_{2211} \right. \right. \nonumber \\
\!&\!-\!&\! \left.\left. f_{1122}-f_{2222} \right)+\left( a_1^c-a_2^c \right)\left( f_{1112}+f_{2212} \right. \right. \nonumber \\
\!&\!+\!&\! \left. \left. f_{1121}+f_{2221} \right) \right] \nonumber \\
V_{S T_0} \!&\!=\!&\! V_{T_0 T_0} + {\cal N}_t^2 \left( a_1^t + a_2^t \right)\left[ \sqrt{2} \left( f_{1111}+f_{1122} \right. \right. \nonumber \\
\!&\!-\!&\! \left. \left. f_{2211}-f_{2222} \right)+\left(a_1^t-a_2^t\right)\left( f_{1211}+ f_{1222} \right. \right. \nonumber \\
\!&\!+\!&\! \left. \left. f_{2111}+f_{2122} \right) \right]\\
V_{SS} \!&\!=\!&\! V_{T_0 T_0}+V_{T_0 S}+V_{S T_0}+{\cal N}_t^2 {\cal N}_c^2 \left( a_1^t+a_2^t \right) \nonumber \\
\!&\! \times \!&\! \left( a_1^c+a_2^c \right)\left[ 2\left( {f_{1111}+f_{2222}-f_{1122}-f_{2211}} \right) \right. \nonumber \\
\!&\!+\!&\! \left.  \sqrt 2 \left( a_1^t-a_2^t \right)\left( f_{1211}+f_{2111}-f_{1222}-f_{2122} \right) \right. \nonumber \\
\!&\!+\!&\! \left. \sqrt 2 \left(a_1^c - a_2^c \right)\left( f_{1112}+f_{1121}-f_{2212}-f_{2221} \right) \right. \nonumber \\
\!&\!+\!&\! \left.  \left(a_1^t-a_2^t\right)\left(a_1^c-a_2^c \right) \right. \nonumber \\
\!&\!\times \!&\! \left.\left( f_{1212}+ f_{2121}+f_{1221}+f_{2121} \right) \right], \nonumber
\end{eqnarray}
where the superscript $t$ ($c$) denotes target (control) qubit, $a_1,a_2$ are the $S(2,0)$, $S(0,2)$ components of the lowest lying hybridized singlet, respectively, and ${\cal N}$ is the normalization.

The Coulomb interaction terms $f_{ijkl}$ are calculated by evaluating the electrostatic energy associated with placing the control qubit charge distribution in the potential $\Phi^t_{ij}$, that is due to the target qubit charge distribution. Denoting  the charge, dipole, and quadrupole electric moments associated with the qubit charge distribution by $q_{ij},{\bf P}_{ij},Q_{ij}$, respectively,
we obtain $f_{ijkl}$ up to and including quadrupole-quadrupole order:
\begin{eqnarray}
f_{ijkl}\!&\! = \!&\! \frac{e^2}{\varepsilon R}+\frac{e}{\varepsilon R^2} \left( P^t_{ij}+P^c_{kl}\right) \nonumber \\
\!&\!-\!&\! \frac{2}{\varepsilon R^3} P^t_{ij} P^c_{kl} + \frac{e}{2\varepsilon R^3} \left(Q^t_{ij(xx)} +Q^c_{kl(xx)} \right) \nonumber \\
\!&-\!&\! \frac{3}{2 \varepsilon R^4} \left( Q^t_{ij(xx)} P^c_{kl} +P^t_{ij} Q^c_{kl(xx)} \right) +\frac{1}{\varepsilon R^5} \label{multipole} \\
\!&\!\times\!&\!  \left[ \frac{1}{6} \sum_{m} Q^t_{ij(mm)} Q^c_{kl(mm)} + \frac{5}{4} Q^t_{ij(xx)} Q^c_{kl(xx)}\! \right] \!. \nonumber
\end{eqnarray}
In deriving Eq.~(\ref{multipole}) we assumed both qubits are aligned with the $x$-axis. The qubit dipole moments are thus in the $x$ direction for both qubits, ${\bf P}_{ij}=\int d{\bf r} {\bf r} \rho_{ij}({\bf r})\hat{x}$. The quadrupole moments, $Q_{ij(mn)}=\int d{\bf r} (3r_m r_n-r^2 \delta_{mm}) \rho_{ij}({\bf r})$, are found to have only diagonal elements ($m=n$).

Explicit expressions for the various multipole moments are given in \cite{Ramon}, and are used to obtain the inter-qubit coupling terms, to quadrupole-quadrupole order with the following nonvanishing contributions:
\begin{eqnarray}
\beta_t\!&\!=\!&\!\beta_t^{21}+\beta_t^{22}+\beta_t^{41}+\beta_t^{24}+ \beta_t^{42}+\beta_t^{44} \label{beta} \\
\gamma\!&\!=\!&\!\gamma^{22}+\gamma^{24}+\gamma^{42}+\gamma^{44} \label{gamma}.
\end{eqnarray}
The first (second) superscript in each term denotes contribution from the particular multipole moment: monopole (1), dipole (2), and quadrupole (4) of the target (control) qubit charge distribution. The effective exchange contribution for the control qubit, $\beta_c$, is found from Eq.~(\ref{beta}) by interchanging the left and right superscripts. Symmetry considerations suggest that for identical qubits, equally biased, one has: $\beta_t^{ij}=\beta_c^{ji}$, and $\gamma^{ij}=\gamma^{ji}$. Note that the dipole moment of a non-biased double dot vanishes, therefore there are no contributions to $\beta_t$ that involve a non-biased target qubit dipole moment. Similarly $\beta_c$ will have no control qubit dipole contributions for non-biased control qubit, and $\gamma$ will have no dipole contributions from either qubits if both are non-biased.

\begin{figure}[tb]
\epsfxsize=0.75\columnwidth
\vspace*{0 cm}
\centerline{\epsffile{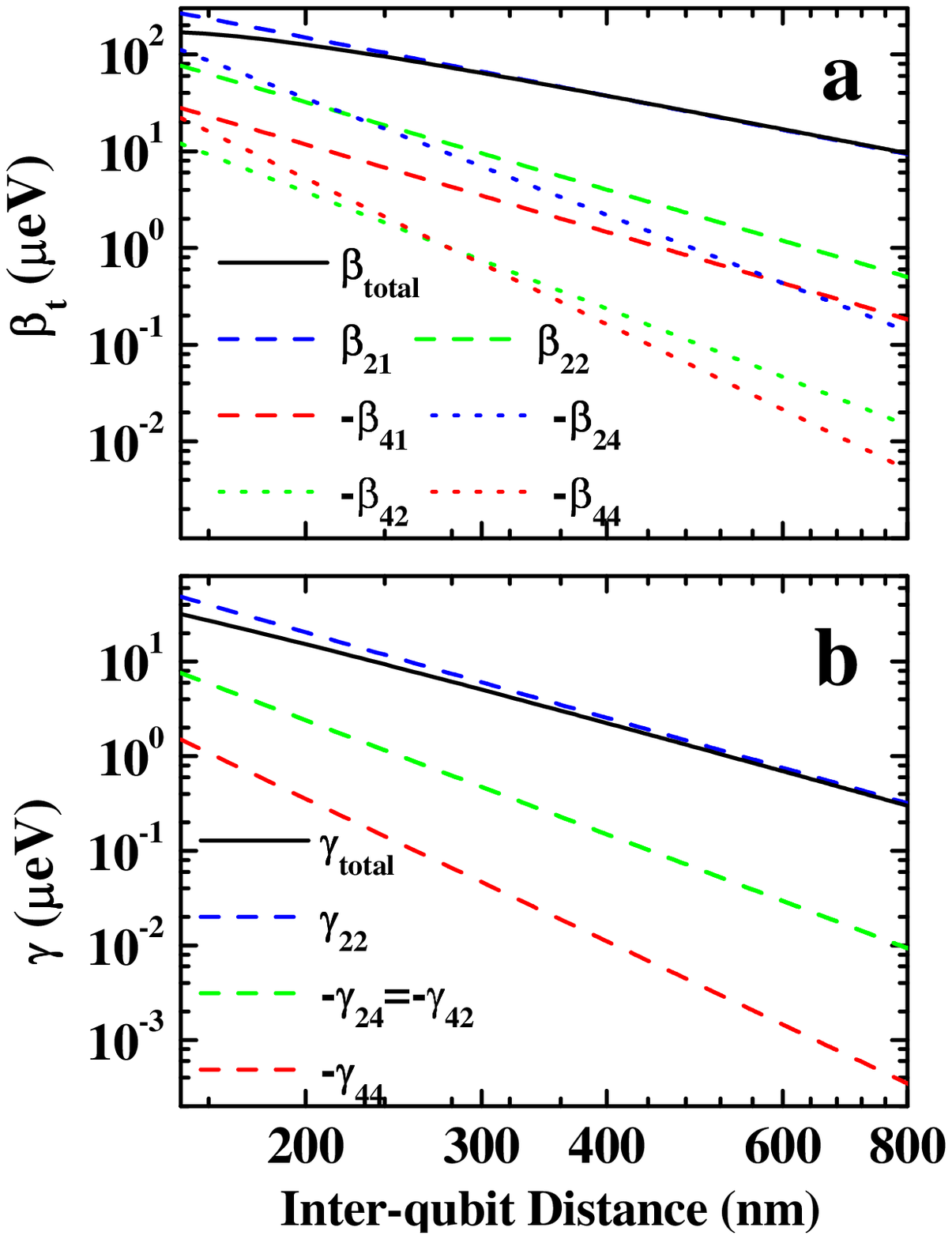}}
\vspace*{-0.1 cm}
\caption{(color online) Multipole expansion contributions to inter-qubit Coulomb coupling vs. inter-qubit distance. Both double dots are biased at the $S(1,1)$-$S(0,2)$ anticrossing point. (a) $\beta$ terms; (b) $\gamma$ terms.}
\label{FigS1}
\end{figure}
Figure \ref{FigS1} shows the dependence of the various multipole terms on the inter-qubit distance for the case of two double-dots equally biased at the singlet anticrossing point. The figure demonstrates the convergence of the multipole expansion in the considered distance range. We conclude that for $R\geqslant 300$ nm, used throughout most of this work, both $\beta$ and $\gamma$ couplings are well approximated by their respective leading contributions.

\subsection*{Inter-qubit dynamics}

The two-qubit Hamiltonian, ${\cal H}={\cal H}_{\rm int}+{\cal H}_t+{\cal H}_c$, is written in the computation basis states: $\{SS,T_0 S,S T_0,T_0 T_0\}$, where the left (right) script denotes the target (control) qubit state, as:
\begin{widetext}
\begin{equation}
{\cal H} = \left( \begin{array}{cccc}
   \frac{{J_t  + J_c }}{2} - \beta_t  - \beta_c  - \gamma  & {\frac{{\delta h_t }}{2}} & {\frac{{\delta h_c }}{2}} & 0  \\
   {\frac{{\delta h_t }}{2}} & { - \frac{{J_t  - J_c }}{2} + \beta_t  - \beta_c  + \gamma } & 0 & {\frac{{\delta h_c }}{2}}  \\
   {\frac{{\delta h_c }}{2}} & 0 & {\frac{{J_t  - J_c }}{2} - \beta_t  + \beta_c  + \gamma } & {\frac{{\delta h_t }}{2}}  \\
   0 & {\frac{{\delta h_c }}{2}} & {\frac{{\delta h_t }}{2}} & { - \frac{{J_t  + J_c }}{2} + \beta_t  + \beta_c  - \gamma }  \\
\end{array} \right).
\label{H}
\end{equation}
\end{widetext}
At the sweet-spot one has: $2\beta_q=J_q$, $q=t,c$. We consider dephasing of the target qubit, initially prepared in the state: $(|S\rangle +|T_0 \rangle)/\sqrt{2}$, by solving for the time evolution of the system of two qubits under the Hamiltonian, Eq.~(\ref{H}), and tracing out the control qubit states. Notice that the term dephasing is used here and in the main text to quantify different time evolutions of $S$ and $T_0$ constituents in a superposition state due to inter-qubit coupling. This is not pure dephasing in the sense of loss of coherence due to entanglement with a bath. For a control field gradient satisfying $\delta h_c \gg \gamma$, which is easily met at the low-bias sweet-spot with a conservative magnetic field gradient of $40$ mT, we find:
\begin{equation}
\rho^t_{ST_0}(t)-\frac{1}{2}=  \frac{2 \gamma^2}{\delta h_c^2} \left( \cos 2\sqrt{ \gamma^2+\left(\frac{\delta h_c}{2}\right)^2} t -1 \right).
\label{rho12b}
\end{equation}
While oscillations are much more rapid ($\sim \delta h_c$) than in the $\delta h_c=0$ case ($\sim 2 \gamma$), their amplitude is very small, thus the qubits are effectively decoupled.

\begin{figure}[tb]
\epsfxsize=0.85\columnwidth
\vspace*{0 cm}
\centerline{\epsffile{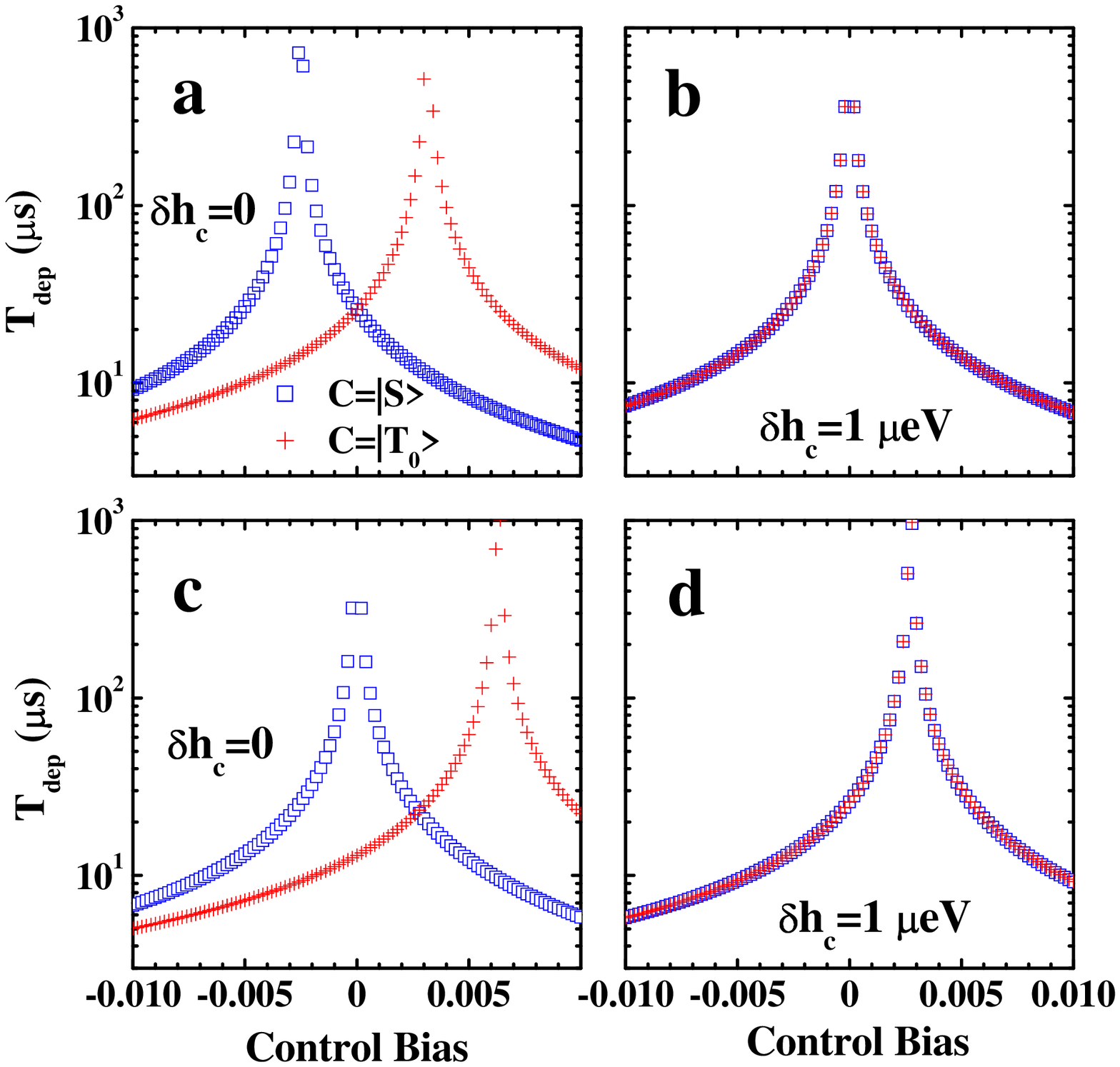}}
\vspace*{-0.2 cm}
\caption{(color online) Dephasing times of target qubit vs. control qubit bias measured from the low-bias sweet spot, at $R=300$ nm. The control qubit is prepared in a singlet (blue squares) or triplet (red plus symbols) state, with magnetic field gradient of: (a,c) $\delta h_c=0$, (b,d) $\delta h_c=1 \mu eV$. In figures (c) and (d) the sweet spot is shifted, satisfying $J_q-2(\beta_q+\gamma)=0$ inboth qubits.}
\label{FigS2}
\end{figure}

To gain better insight to the qubit decoupling bias points and their robustness against bias errors we plot in Figs.~\ref{FigS2}(a,b) target qubit dephasing times vs. the control qubit bias, measured from the $J_c=2\beta_c$, low-bias position. We define dephasing time as the time it takes $\rho^t_{S T_0}(t)$ to drop to 50\% of its original value of $1/2$, due to coupling with the control qubit. In general, we note that the effects of the control qubit on target dephasing are minimal at a relatively wide control bias region around the sweet spot. This should be contrasted with a much higher sensitivity to bias errors at the high-bias position.

When $\delta h_c =0$, target dephasing at the sweet spot is identical for both control states, but Fig.~\ref{FigS2}(a) shows that slight detuning of the control bias results in complete qubit decoupling when the control is in either singlet or triplet state. To explain this behavior we notice that at bias positions satisfying $J_q-2(\beta_q \pm \gamma)=0$ in both qubits, Eq.~(\ref{H}) leads to: $\exp(-i{\cal H} t)={\rm diag} (1,1,e^{-2i\gamma t},e^{2i\gamma t})$ for the plus sign, and  $\exp(-i{\cal H} t)={\rm diag} (e^{-2i\gamma t},e^{2i\gamma t},1,1)$ for the minus sign. As a result, the qubits are decoupled when the control is in a singlet (triplet) state for the plus (minus) sign, and the target qubit dephases as $\rho^t_{ST_0}(t)=\frac{1}{2} \cos (4 \gamma t)$ when the control is in the other state. Since $\beta_c$ grows faster than $J_c$ with $\epsilon_c$, the intersection $J_c=2(\beta_c + \gamma)$ occurs at a negative bias, and similarly $J_c=2(\beta_c - \gamma)$ occurs at a positive bias relative to the sweet spot position (see figure \ref{Fig1}(b)). These considerations are numerically confirmed by correcting the sweet spot to satisfy $J_q-2(\beta_q + \gamma)=0$ for $q=t,c$. Fig. \ref{FigS2}(c) shows qubit decoupling at this $\gamma$-corrected sweet spot position for singlet control state. When $\delta h_c \gg \gamma$, the qubits decouple at the original sweet spot (with errors of order $\sim (\gamma/\delta h_c)^2$) according to Eq.~(\ref{rho12b}), for both control states, thus shifting the sweet spot bias results in a similar shift in the decoupling bias position, as seen in Fig.~\ref{FigS2}(d).

Finally, errors in $X$-rotations performed at the $\tilde{J}_t=0$ bias position can be evaluated, using Eq.~(\ref{H}). Considering a $\pi$-flip on a target qubit, initially prepared in a singlet state, we take $\tau=\pi/\delta h_t$ and find the leading contribution in a $\gamma/\delta h$ expansion. At the end of the pulse, the gate error, given by the singlet probability is found to be:
\begin{equation}
P_S(\tau) \approx \left( \frac{2 \gamma \cos \left(\frac{\pi}{2} \delta h_c/\delta h_t \right)}{\delta h_t (1-\delta h_c^2/\delta h_t^2)} \right)^2,
\label{Xerr}
\end{equation}
irrespective of the control qubit state. When $\delta h_t=\delta h_c =\delta h $ Eq.~(\ref{Xerr}) reduces to $P_S(\tau)=(\pi \gamma/2 \delta h)^2$, given in the main text. Notice, that when $\delta h_c=n \delta h_t$, where $n$ is an odd integer larger than 1, the leading contribution, Eq.~(\ref{Xerr}), vanishes and we get $X$-gate error of the order $(\gamma/\delta h_t)^4$.

\subsection*{Constructing $Z$-rotations with finite $\delta h$}

Here we use a variant of the bias sequence generating $X$-rotations that was proposed in \cite{HanBur}. Our cycle consists of two working points: (i) the low-bias symmetric sweet-spot $\tilde{J}_t =0$, where rotations are about the $X$ axis, and (ii) a bias position slightly above the $\tilde{J}_t =0$ point, where $\tilde{J}_t \gtrsim \delta h_t$ and the rotation is about a $\theta$-tilted axis where $\theta=\tan^{-1} (\delta h_t/\tilde{J}_t)$. When $|\theta| \leq \pi/4$, a $Z$-rotation by an arbitrary angle $\alpha$ can be generated by the cycle:
\begin{equation}
U_z (\alpha)=U_\theta (\chi) U_x (\phi) U_\theta (\chi),
\label{txt}
\end{equation}
where $U_\theta (\chi)$ and $U_x (\phi)$ are rotation matrices about the $\theta$-tilted  and $X$ axes, respectively, given by:
\begin{eqnarray}
U_\theta (\chi) \!&\!=\!&\!
\left( \begin{array}{cc}
\cos \frac{\chi}{2}-i\sin \frac{\chi}{2} \cos \theta & -i\sin \frac{\chi}{2} \sin \theta \\
-i\sin \frac{\chi}{2} \sin \theta & \cos \frac{\chi}{2}+i\sin \frac{\chi}{2} \cos \theta
\end{array} \right) \nonumber \\
U_x (\phi) \!&\!=\!&\!
\left( \begin{array}{cc}
\cos \frac{\phi}{2} & i\sin \frac{\phi}{2} \\
i\sin \frac{\phi}{2} & \cos \frac{\phi}{2}
\end{array} \right).
\end{eqnarray}
The angles $\chi$ and $\phi$ are functions of $\theta$ and $\alpha$, and are found to be similar (but not identical) to those given in \cite{HanBur}:
\begin{eqnarray}
\chi \!&\!=\!&\! \arccos \frac{-\cos\frac{\alpha}{2} \sqrt{1-\tan^2 \theta \sin^2 \frac{\alpha}{2}}-\sin^2 \frac{\alpha}{2} \sin^2 \theta}{\cos^2 \frac{\alpha}{2} +\cos^2 \theta \sin^2\frac{\alpha}{2}} \\
\phi\!&\!=\!&\! 2 \arctan \frac{\pm \sin \chi \sin \theta}{\cos^2 \frac{\chi}{2} +\cos 2 \theta \sin^2 \frac{\chi}{2}},
\label{phi}
\end{eqnarray}
where the plus (minus) sign in Eq.~(\ref{phi}) corresponds to positive (negative) $\theta$.
In particular, a $\pi$-rotation about the $Z$ axis is obtained with:
\begin{equation}
\chi=\arccos (-\tan^2 \theta); \hspace{0.5 cm} \phi=\pm2\arctan \frac{\sin \theta}{\sqrt{\cos 2\theta}}.
\label{chiphi}
\end{equation}

We consider a $\pi$-rotation about the $Z$ axis by taking the second working point to satisfy $\theta=\pi/4$ so that the bias sweep from the low-bias sweet-spot is minimized. The pulse durations are set to: $\tau_\theta=\pi/\sqrt{2} \delta h_t$, and $\tau_x =\pi/\delta h_t$, at the $\theta$-tilted point and low-bias sweet-spot, respectively. To simulate this cycle we discretize bias and time steps, so that the actual switching time between the working points is taken into account. Bias switching should be nonadiabatic with respect to nuclear mixing, but slow as compared with the tunnel splitting of the hybridized singlet states. Bias sweep rates $\lesssim 2$ mV/ns, comparable to experimental values \cite{Petta}, do not contribute appreciably to gate errors presented in this work, and can be further corrected by pulse design.

\begin{figure}[tb]
\epsfxsize=0.9\columnwidth
\vspace*{3.1 cm}
\centerline{\epsffile{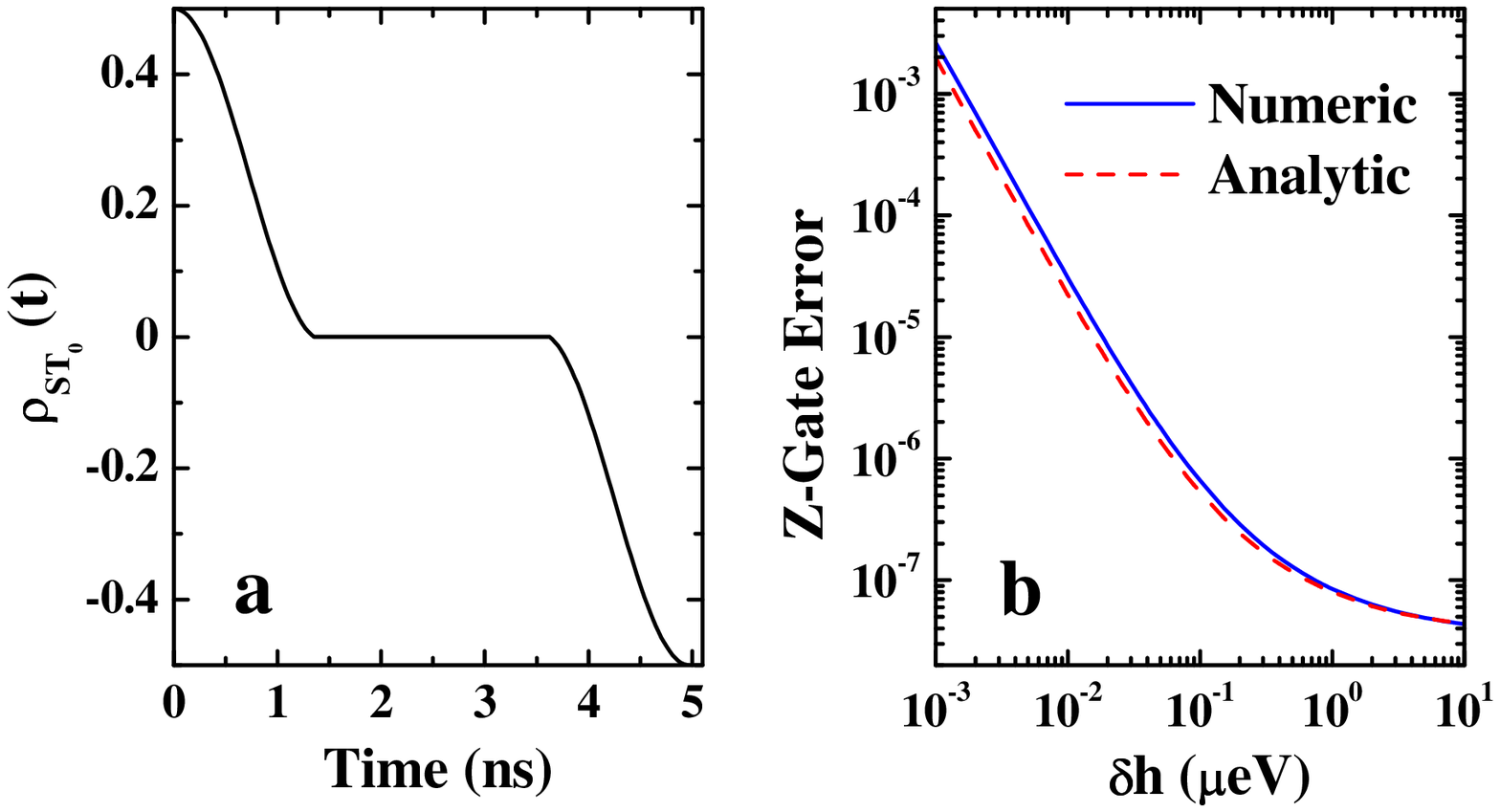}}
\vspace*{-3.6 cm}
\caption{(color online) (a) $\pi$-rotation about the $Z$ axis using a three-pulse sequence. $\delta h_t=1 \mu eV$ and $R=300$ nm. The first and third rotations are performed at a bias position satisfying $\tilde{J}_t=\delta h_c$ ($\theta =\pi/4$); (b) Gate error as a function of $\delta h$, evaluated from numerical simulation (solid blue line) and from Eq.~(\ref{Zerr}) (dashed red line).}
\label{FigS3}
\end{figure}

Fig.~\ref{FigS3}(a) depicts the target off-diagonal density matrix element during a $\pi$-rotation about the $Z$ axis from $(|S \rangle+|T_0 \rangle)/\sqrt{2} \rightarrow (|S \rangle-|T_0 \rangle)/\sqrt{2}$, using the sequence in Eq.~(\ref{txt}). Taking $\delta h_t =\delta h_c =\delta h$ we use Eq.~(\ref{H}) to find the deviation of the the target off diagonal density matrix element from $-1/2$ at the end of the cycle. The leading error terms in a $\gamma_x/\delta h$, $\gamma_\theta/\delta h$ expansion, where $\gamma_x$ ($\gamma_\theta$) is evaluated at the sweet-spot ($\theta$-tilted) position, are found to be:
\begin{eqnarray}
\rho_{S T_0}^t (2\tau_\theta+ \tau_x)  \!&\!+\!&\! \frac{1}{2} = \frac{1}{\delta h^2} \left[ \left(\frac{\pi}{2} \gamma_x \sin \frac{\pi}{\sqrt{2}}+2 \gamma_\theta \right)^2 \right. \nonumber \\
\!&\!+\!&\! \left.  \left( \frac{\pi}{2} \gamma_x \cos \frac{\pi}{\sqrt{2}}+ 2\gamma_\theta \sin \frac{\pi}{\sqrt{2}} \right)^2 \right]\! ,
\label{Zerr}
\end{eqnarray}
and are compared with the numerical simulations in Fig.~\ref{FigS3}(b). The proposed sequence, operating around the low-bias sweet-spot, is shown to generate high fidelity $Z$-rotations at a reasonably short gate time.

\subsection*{Alternative CNOT gate}

For demonstration purpose an effective CNOT gate for an initial target qubit in either $|S\rangle$ or $|T_0 \rangle$ can be implemented by biasing both qubits to the point $J_q-2(\beta_q-\gamma)=0$, $q=t,c$, close to the high-bias symmetric sweet-spot discussed in the main text. At this bias, the diagonal part of the two-qubit hamiltonian is $(-3\gamma, \gamma, \gamma, \gamma)$, and $\gamma \gg \delta h \lesssim 1 \mu eV$. Thus, the target state will effectively rotate about the $Z$-axis when the control is in $| S\rangle$ and about the $X$-axis when the control is in $|T_0 \rangle$. Applying this pulse for $\tau=\pi/ \delta h$ to a two-qubit system, with the target initially in either of the computational states (but not in a superposition of the two), results in a CNOT gate, with gate time inversely proportional to $\delta h$. The degeneracy in the Hamiltonian at the above bias point induces gate errors that scale linearly with $\delta h/\gamma$, and therefore are rather large for $R=300$ nm ($\gamma=20 \mu eV$). A better bias position is the slightly asymmetric point: $J_t-2(\beta_t-\gamma)=0$, $J_c-2\beta_c=0$, at which the diagonal part of the two-qubit hamiltonian is $(-2\gamma, 2\gamma, 0, 0)$. At this position gate errors are found to scale as $(\delta h/\gamma)^2$, and are small over a wide range of $\delta h$ values.

CNOT gate can also be implemented using the scheme proposed in \cite{HanBur}, but this scheme requires biasing the two double dots in opposite directions, so that the configuration: $S(0,2)-S(2,0)$ is obtained. This bias position is very far from the low-bias symmetric point that is proposed in the current work for single qubit operations, and could be experimentally challenging.

\subsection*{Effect of TLF number on qubit dephasing}

\begin{figure}[tb]
\epsfxsize=0.9\columnwidth
\vspace*{0.9 cm}
\centerline{\epsffile{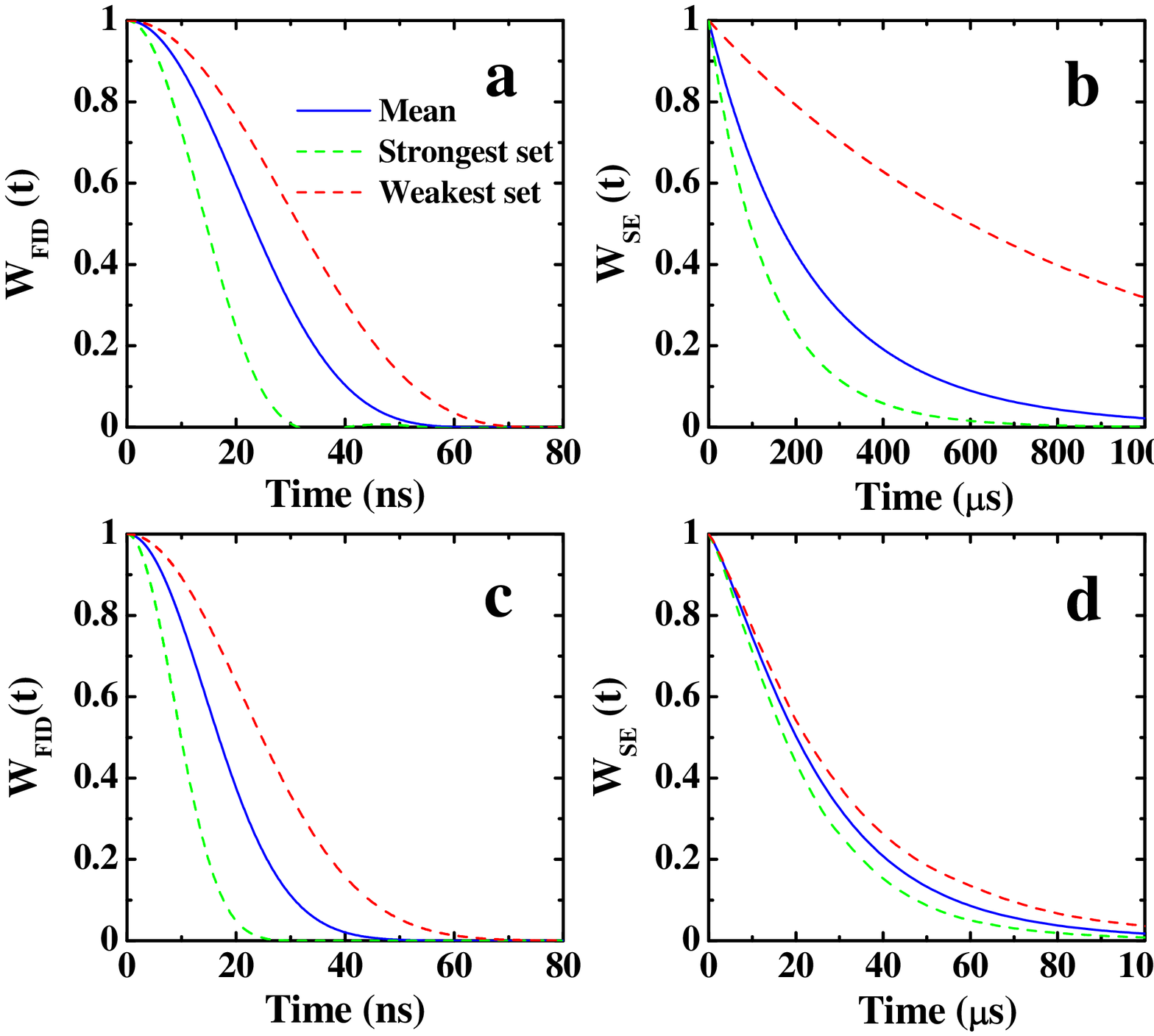}}
\vspace*{-1.1 cm}
\caption{(color online) Two-spin qubit dephasing due an ensemble of TLFs, for two device sizes. Figs.~(a) and (b) show results for 12 TLFs, and Figs.~(c) and (d) show results for 112 TLFs.
Two qubit manipulation protocols are considered: free induction decay in Figs.~(a) and (c), and spin echo in Figs.~(b) and (d). In all figures the solid blue lines correspond to an average over 10,000
sets of TLFs with random locations, orientations, and switching rates, within the parameter ranges specified in the main text. The dashed green and red lines mark the distribution widths of the signal decays.}
\label{FigS4}
\end{figure}

Here we demonstrate the effect of the device size (and hence the number of charge traps) on qubit dephasing and sample-to-sample variability. Figs.~\ref{FigS4} (a) and (b) show calculated FID and SE dephasing, respectively, for a device of area $0.75 \mu m^2$ ($n_\beta =3$, $n_\gamma =9$), considered in the main text. The solid blue lines depict average decay over 10,000 random TLF sets, and the dashed green (red) lines correspond to maximally (minimally) affecting sets. The results show a large sample-to-sample variability, in particular for the SE decay, which is governed by the wide switching rate distribution, 1 ms$<\Gamma^{-1}<$1 s. Figs.~\ref{FigS4} (c) and (d) show FID and SE dephasing, respectively, for a device of area $6 \mu m^2$ ($n_\beta =28$, $n_\gamma =84$). The FID dephasing time, $T_2^* =17$ ns, and variability are hardly changed with respect to the smaller device. In contrast, the SE decay shows an order-of-magnitude reduction in dephasing time, $T_2^{\rm SE} =20 \mu s$, and much more reproducible decay, suggesting that for $n_{\rm trap}\gtrsim 100$ results become self averaging.

We note that in both qubit manipulation protocols considered here, the noise generated by the charge environment is non-Gaussian, and therefore cannot be described in terms of its spectral function alone. For larger devices, where the number of fluctuators is of the order of 100, the distribution of coupling strengths and switching rates can be averaged, and the resulting noise is dominated by a large number of weakly coupled TLFs, demonstrating characteristics of $1/f$ spectrum. The dependence of qubit dephasing on TLF parameter distributions, qubit characteristics, and  magnetic field, and its scaling with the number of qubits, as well as qubit dephasing under multiple-pulse dynamical decoupling sequences, will be presented elsewhere \cite{Ramon2}.

\end{document}